\documentclass[12pt,authoryear]{elsarticle}

\usepackage{amssymb}
\usepackage{amsmath}

\usepackage{hyperref}
\usepackage[locale=UK,
number-mode=math,
unit-mode=text,
per-mode=symbol,
number-unit-product=\ ,
retain-zero-exponent=false]{siunitx}[=v2]

\usepackage{mathtools} 
\usepackage{xstring}						
\usepackage{soul}							
\usepackage{cleveref}					
\usepackage{xcolor}
\usepackage{xspace}
\usepackage{enumitem}
\setlist{noitemsep} 
\usepackage{pdfpages}

\definecolor{epflblue}{RGB}{31,55,91}
\definecolor{epflred}{RGB}{255,0,0}
\definecolor{groseille}{RGB}{181,31,31}
\definecolor{leman}{RGB}{0,167,159}
\definecolor{canard}{RGB}{0,116,128}
\definecolor{ardoise}{RGB}{65,61,58}
\definecolor{perle}{RGB}{202,199,199}
\definecolor{taupe}{RGB}{69, 58, 76}
\definecolor{montrose}{RGB}{242, 151, 105}
\definecolor{vertdeau}{RGB}{193, 220, 175}
\definecolor{rose}{RGB}{236, 110, 155}
\definecolor{acier}{RGB}{79, 142, 203}
\definecolor{souffre}{RGB}{251, 237, 102}
\definecolor{carotte}{RGB}{235, 102, 8}
\definecolor{zinzolin}{RGB}{92, 36, 130}
\definecolor{chartreuse}{RGB}{127, 255, 0}
\definecolor{marron}{RGB}{91, 52, 40}
\definecolor{white}{rgb}{1.000,1.000,1.000}
\definecolor{snow}{rgb}{1.000,0.979,0.979}
\definecolor{honeydew}{rgb}{0.938,1.000,0.938}
\definecolor{mintcream}{rgb}{0.958,1.000,0.979}
\definecolor{azure}{rgb}{0.938,1.000,1.000}
\definecolor{aliceblue}{rgb}{0.938,0.971,1.000}
\definecolor{ghostwhite}{rgb}{0.971,0.971,1.000}
\definecolor{whitesmoke}{rgb}{0.958,0.958,0.958}
\definecolor{seashell}{rgb}{1.000,0.958,0.930}
\definecolor{beige}{rgb}{0.958,0.958,0.859}
\definecolor{oldlace}{rgb}{0.992,0.958,0.898}
\definecolor{floralwhite}{rgb}{1.000,0.979,0.938}
\definecolor{ivory}{rgb}{1.000,1.000,0.938}
\definecolor{antiquewhite}{rgb}{0.979,0.918,0.840}
\definecolor{linen}{rgb}{0.979,0.938,0.898}
\definecolor{lavenderblush}{rgb}{1.000,0.938,0.958}
\definecolor{mistyrose}{rgb}{1.000,0.891,0.879}
\definecolor{gray}{rgb}{0.500,0.500,0.500}
\definecolor{gainsboro}{rgb}{0.859,0.859,0.859}
\definecolor{lightgray}{rgb}{0.824,0.824,0.824}
\definecolor{silver}{rgb}{0.750,0.750,0.750}
\definecolor{darkgray}{rgb}{0.660,0.660,0.660}
\definecolor{dimgray}{rgb}{0.410,0.410,0.410}
\definecolor{lightslategray}{rgb}{0.465,0.531,0.598}
\definecolor{slategray}{rgb}{0.438,0.500,0.562}
\definecolor{darkslategray}{rgb}{0.184,0.309,0.309}
\definecolor{black}{rgb}{0.000,0.000,0.000}
\definecolor{red}{rgb}{1.000,0.000,0.000}
\definecolor{lightsalmon}{rgb}{1.000,0.625,0.477}
\definecolor{salmon}{rgb}{0.979,0.500,0.445}
\definecolor{darksalmon}{rgb}{0.910,0.586,0.477}
\definecolor{lightcoral}{rgb}{0.938,0.500,0.500}
\definecolor{indianred}{rgb}{0.801,0.359,0.359}
\definecolor{crimson}{rgb}{0.859,0.078,0.234}
\definecolor{firebrick}{rgb}{0.695,0.133,0.133}
\definecolor{darkred}{rgb}{0.543,0.000,0.000}
\definecolor{pink}{rgb}{1.000,0.750,0.793}
\definecolor{lightpink}{rgb}{1.000,0.711,0.754}
\definecolor{hotpink}{rgb}{1.000,0.410,0.703}
\definecolor{deeppink}{rgb}{1.000,0.078,0.574}
\definecolor{palevioletred}{rgb}{0.855,0.438,0.574}
\definecolor{mediumvioletred}{rgb}{0.777,0.082,0.520}
\definecolor{orange}{rgb}{1.000,0.645,0.000}
\definecolor{darkorange}{rgb}{1.000,0.547,0.000}
\definecolor{coral}{rgb}{1.000,0.496,0.312}
\definecolor{tomato}{rgb}{1.000,0.387,0.277}
\definecolor{orangered}{rgb}{1.000,0.270,0.000}
\definecolor{yellow}{rgb}{1.000,1.000,0.000}
\definecolor{lightyellow}{rgb}{1.000,1.000,0.875}
\definecolor{lemonchiffon}{rgb}{1.000,0.979,0.801}
\definecolor{lightgoldenrodyellow}{rgb}{0.979,0.979,0.820}
\definecolor{papayawhip}{rgb}{1.000,0.934,0.832}
\definecolor{moccasin}{rgb}{1.000,0.891,0.707}
\definecolor{peachpuff}{rgb}{1.000,0.852,0.723}
\definecolor{palegoldenrod}{rgb}{0.930,0.906,0.664}
\definecolor{khaki}{rgb}{0.938,0.898,0.547}
\definecolor{darkkhaki}{rgb}{0.738,0.715,0.418}
\definecolor{gold}{rgb}{1.000,0.840,0.000}
\definecolor{brown}{rgb}{0.645,0.164,0.164}
\definecolor{cornsilk}{rgb}{1.000,0.971,0.859}
\definecolor{blanchedalmond}{rgb}{1.000,0.918,0.801}
\definecolor{bisque}{rgb}{1.000,0.891,0.766}
\definecolor{navajowhite}{rgb}{1.000,0.867,0.676}
\definecolor{wheat}{rgb}{0.958,0.867,0.699}
\definecolor{burlywood}{rgb}{0.867,0.719,0.527}
\definecolor{tan}{rgb}{0.820,0.703,0.547}
\definecolor{rosybrown}{rgb}{0.734,0.559,0.559}
\definecolor{sandybrown}{rgb}{0.954,0.641,0.375}
\definecolor{goldenrod}{rgb}{0.852,0.645,0.125}
\definecolor{darkgoldenrod}{rgb}{0.719,0.523,0.043}
\definecolor{peru}{rgb}{0.801,0.520,0.246}
\definecolor{chocolate}{rgb}{0.820,0.410,0.117}
\definecolor{saddlebrown}{rgb}{0.543,0.270,0.074}
\definecolor{sienna}{rgb}{0.625,0.320,0.176}
\definecolor{maroon}{rgb}{0.500,0.000,0.000}
\definecolor{green}{rgb}{0.000,0.500,0.000}
\definecolor{palegreen}{rgb}{0.594,0.983,0.594}
\definecolor{lightgreen}{rgb}{0.562,0.930,0.562}
\definecolor{yellowgreen}{rgb}{0.602,0.801,0.195}
\definecolor{greenyellow}{rgb}{0.676,1.000,0.184}
\definecolor{chartreuse}{rgb}{0.496,1.000,0.000}
\definecolor{lawngreen}{rgb}{0.484,0.988,0.000}
\definecolor{lime}{rgb}{0.000,1.000,0.000}
\definecolor{limegreen}{rgb}{0.195,0.801,0.195}
\definecolor{mediumspringgreen}{rgb}{0.000,0.979,0.602}
\definecolor{springgreen}{rgb}{0.000,1.000,0.496}
\definecolor{mediumaquamarine}{rgb}{0.398,0.801,0.664}
\definecolor{aquamarine}{rgb}{0.496,1.000,0.828}
\definecolor{lightseagreen}{rgb}{0.125,0.695,0.664}
\definecolor{mediumseagreen}{rgb}{0.234,0.699,0.441}
\definecolor{seagreen}{rgb}{0.180,0.543,0.340}
\definecolor{darkseagreen}{rgb}{0.559,0.734,0.559}
\definecolor{forestgreen}{rgb}{0.133,0.543,0.133}
\definecolor{darkgreen}{rgb}{0.000,0.391,0.000}
\definecolor{olivedrab}{rgb}{0.418,0.555,0.137}
\definecolor{olive}{rgb}{0.500,0.500,0.000}
\definecolor{darkolivegreen}{rgb}{0.332,0.418,0.184}
\definecolor{teal}{rgb}{0.000,0.500,0.500}
\definecolor{blue}{rgb}{0.000,0.000,1.000}
\definecolor{lightblue}{rgb}{0.676,0.844,0.898}
\definecolor{powderblue}{rgb}{0.688,0.875,0.898}
\definecolor{paleturquoise}{rgb}{0.684,0.930,0.930}
\definecolor{turquoise}{rgb}{0.250,0.875,0.812}
\definecolor{mediumturquoise}{rgb}{0.281,0.816,0.797}
\definecolor{darkturquoise}{rgb}{0.000,0.805,0.816}
\definecolor{lightcyan}{rgb}{0.875,1.000,1.000}
\definecolor{cyan}{rgb}{0.000,1.000,1.000}
\definecolor{aqua}{rgb}{0.000,1.000,1.000}
\definecolor{darkcyan}{rgb}{0.000,0.543,0.543}
\definecolor{cadetblue}{rgb}{0.371,0.617,0.625}
\definecolor{lightsteelblue}{rgb}{0.688,0.766,0.867}
\definecolor{steelblue}{rgb}{0.273,0.508,0.703}
\definecolor{lightskyblue}{rgb}{0.527,0.805,0.979}
\definecolor{skyblue}{rgb}{0.527,0.805,0.918}
\definecolor{deepskyblue}{rgb}{0.000,0.746,1.000}
\definecolor{dodgerblue}{rgb}{0.117,0.562,1.000}
\definecolor{cornflowerblue}{rgb}{0.391,0.582,0.926}
\definecolor{royalblue}{rgb}{0.254,0.410,0.879}
\definecolor{mediumblue}{rgb}{0.000,0.000,0.801}
\definecolor{darkblue}{rgb}{0.000,0.000,0.543}
\definecolor{navy}{rgb}{0.000,0.000,0.500}
\definecolor{midnightblue}{rgb}{0.098,0.098,0.438}
\definecolor{purple}{rgb}{0.500,0.000,0.500}
\definecolor{lavender}{rgb}{0.898,0.898,0.979}
\definecolor{thistle}{rgb}{0.844,0.746,0.844}
\definecolor{plum}{rgb}{0.863,0.625,0.863}
\definecolor{violet}{rgb}{0.930,0.508,0.930}
\definecolor{orchid}{rgb}{0.852,0.438,0.836}
\definecolor{fuchsia}{rgb}{1.000,0.000,1.000}
\definecolor{magenta}{rgb}{1.000,0.000,1.000}
\definecolor{mediumorchid}{rgb}{0.727,0.332,0.824}
\definecolor{mediumpurple}{rgb}{0.574,0.438,0.855}
\definecolor{amethyst}{rgb}{0.598,0.398,0.797}
\definecolor{blueviolet}{rgb}{0.539,0.168,0.883}
\definecolor{darkviolet}{rgb}{0.578,0.000,0.824}
\definecolor{darkorchid}{rgb}{0.598,0.195,0.797}
\definecolor{darkmagenta}{rgb}{0.543,0.000,0.543}
\definecolor{slateblue}{rgb}{0.414,0.352,0.801}
\definecolor{darkslateblue}{rgb}{0.281,0.238,0.543}
\definecolor{mediumslateblue}{rgb}{0.480,0.406,0.930}
\definecolor{indigo}{rgb}{0.293,0.000,0.508}
\definecolor{grey}{rgb}{0.500,0.500,0.500}
\definecolor{lightgrey}{rgb}{0.824,0.824,0.824}
\definecolor{darkgrey}{rgb}{0.660,0.660,0.660}
\definecolor{dimgrey}{rgb}{0.410,0.410,0.410}
\definecolor{lightslategrey}{rgb}{0.465,0.531,0.598}
\definecolor{slategrey}{rgb}{0.438,0.500,0.562}
\definecolor{darkslategrey}{rgb}{0.184,0.309,0.309}
\definecolor{rosa}{rgb}{1,0.5,0.5}
\definecolor{parula-1}{rgb}{0.2081,0.1663,0.5292}
\definecolor{parula-2}{rgb}{0.0146,0.3845,0.8813}
\definecolor{parula-3}{rgb}{0.0795,0.5159,0.8328}
\definecolor{parula-4}{rgb}{0.0228,0.6492,0.7823}
\definecolor{parula-5}{rgb}{0.1986,0.7214,0.6310}
\definecolor{parula-6}{rgb}{0.5456,0.7490,0.4597}
\definecolor{parula-7}{rgb}{0.8266,0.7320,0.3464}
\definecolor{parula-8}{rgb}{0.9948,0.7886,0.1943}
\definecolor{parula-9}{rgb}{0.9763,0.9831,0.0538}

\newcommand{\Rey}{\textit{Re}\xspace}
\newcommand{\Ca}{\textit{Ca}\xspace}

\graphicspath{{./figures}}

\usepackage{lineno}

\journal{Journal of Fluids and Structures}

\begin{document}
	
	\begin{frontmatter}
		
		\title{Coupled poro-elastic behavior of hyper-elastic membranes}
		
		\author[label1]{Alexander Gehrke}
		\author[label1]{Zoe King}
		\author{Kenneth S. Breuer\corref{cor1}\fnref{label1}}
		
		\affiliation[label1]{organization={School of Engineering, Brown University}, 
			addressline={182 Hope Street}, 
			city={Providence},
			postcode={02912}, 
			state={Rhode Island},
			country={United States of America}}
		\cortext[cor1]{Corresponding author. E-mail address: \href{mailto:kenneth_breuer@brown.edu}{kenneth\_breuer@brown.edu}}
		
		\begin{abstract}
			This study investigates the coupled deformation and flow behavior of thin, hyper-elastic, porous membranes subjected to pressure loading. Using bulge test experiments, optical deformation measurements, and flow rate characterization, we analyze the structural and fluid dynamic responses of membranes with varying material stiffness and porosity patterns. A two-parameter Gent model accurately captures the hyper-elastic deformation, and local stretch analysis reveals non-uniform stretch distributions across the membrane. We find that membrane deformation is primarily governed by material stiffness and pressure, independent of porosity. Pore diameter scales linearly with local stretch, leading to a radial gradient of increasing pore size toward the membrane center. Flow rate scaling is characterized using a discharge coefficient, which accounts for both pore area expansion and pressure losses. Together, these results establish a unified framework that links structural deformation and flow performance in flexible porous membranes, providing robust scaling laws for the design of adaptive, bio-inspired flow-regulating systems.
		\end{abstract}
		
		
	\end{frontmatter}
	

\section{Introduction}
%
%
Poro-elastic materials combine flexibility with porosity and are widespread in both natural and engineered systems across a large range of length scales.
In nature, poro-elastic structures are found in plant components such as leaves and seeds, or in trees, which use their porosity to reduce structural loads \citep{vogel_drag_1984, vogel_drag_1989} or to enable efficient seed dispersal \citep{cummins_separated_2018}.
Similarly, the bristled wings of tiny insects and bird feathers employ porosity to enhance aerodynamic performance by being lightweight and resilient, thus enabling both efficient \citep{eberle_fluidstructure_2014, santhanakrishnan_clap_2014, kolomenskiy_aerodynamic_2020, jiang_bristled-wing_2022} and silent flight \citep{jaworski_aerodynamic_2013, jaworski_aeroacoustics_2020}.
Agglomerations of individual non-porous bundles can act as poro-elastic terrestrial \citep{brunet_turbulent_2020} and aquatic \citep{nepf_flow_2012} canopies, even when the individual blades or leaves are non-porous.

In engineering, poro-elastic materials are critical for applications ranging from rock mechanics to tissue engineering and medical devices.
For instance, poro-elasticity plays a crucial role in the mechanics of porous rock sediments, where it is used to model subsurface fluid flow \citep{detournay_comprehensive_1993, steeb_mechanics_2019}.
In the medical field, the study of the poro-elastic properties of bones \citep{ehret_inverse_2017} and living tissues \citep{nia_poroelasticity_2011, malandrino_poroelasticity_2019} has informed the development of implants and other medical devices \citep{roche_bioinspired_2014}.
Moreover, soft elastomeric materials are increasingly used in applications like soft robotics \citep{shian_dielectric_2015, whitesides_soft_2018, jones_bubble_2021}, and bio-inspired engineering \citep{song_aeromechanics_2008, gehrke_aeroelastic_2022, gehrke_highly_2025}, due to their ability to undergo large deformations while maintaining their structural integrity \citep{michel_comparison_2010}.

Despite significant progress in the modeling of fluid-structure interactions \citep{dowell_modeling_2001, venkataraman_minimal_2014}, the role of poro-elastic interactions, especially in the context of inertia, remains underexplored \citep{tiomkin_review_2021}.
This is particularly evident in aerodynamics, where porosity can significantly impact the performance and efficiency of various structures, such as parachutes \citep{heinrich_stability_1971, kim_2d_2006, kanai_compressible-flow_2019}, porous airfoils \citep{geyer_measurement_2010, aldheeb_aerodynamics_2018}, and sails \citep{murata_aerodynamic_1989}.
While there is a particular gap in experimental poro-elastic studies, significant effort has been made in theoretical modeling \citep{iosilevskii_aerodynamics_2011, iosilevskii_aerodynamics_2013, baddoo_unsteady_2021}, and in individual studies of the effect of compliant membranes \citep{mathai_shape-morphing_2023} and porous screens \citep{pinker_pressure_1967, castro_wake_1971, koo_fluid_1973, cummins_effect_2017}. \\

This paper focuses on the experimental characterization of large deformations and flow rates through highly flexible poro-elastic membranes under varying pressure loadings.
Building on the work of \citet{das_nonlinear_2020}, we fabricate membranes with different material strengths and porosity patterns, employing a fabrication process that allows precise control of mechanical properties and pore distribution.
By subjecting these membranes to controlled pressure conditions, we gain new insights into the complex fluid-structure interactions governing their response.
Our methods include optical deformation measurements, local stretch analysis to capture pore size variation along the membrane, and flow rate measurements through the open pores to characterize the discharge.

We integrate a two-parameter Gent model for the deformation scaling and extend it to determine the pore-size evolution across the membrane.
Combining these models enables us to predict the three-dimensional shape of the membranes along with their pore size distribution under varying pressure loadings.
By defining a pore discharge coefficient, we establish a scaling ratio that captures the relationship between flow rates across membranes with different material strengths and porosity patterns.

This study advances our understanding of poro-elastic membranes and provides practical guidelines for applying porous fluid-structure interaction systems, e.g. in bio-inspired aerodynamic designs.
Our research contributes to the broader understanding of poro-elastic materials and their applications in both natural and engineered systems.
The methodologies and findings presented here are expected to pave the way for future research and innovations in utilizing these materials for enhanced performance and efficiency.
\section{Experimental Methods}
The primary objective of this study is to characterize the effects of poro-elasticity on the deformation and flow rate through thin, highly compliant membranes.
To this end, membranes of varying material strengths and porosity rates are fabricated and characterized using a bulge test pressure chamber.
In the following sections, we outline the different porosity patterns employed, the membrane fabrication process, and introduce the experimental test stand with all its components.
\subsection{Poro-elastic membranes} \label{sec:method_membranes}
\begin{figure*}
	\centering
	\includegraphics[width=\textwidth]{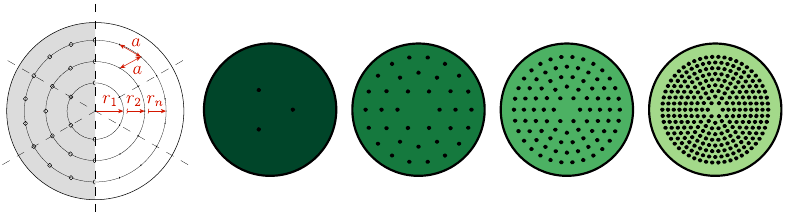}
	\caption[]{
		Porosity definition for the radial six-spoke pattern and examples of patterns with 1, 3, 5, and 9 layers of pores leading to initial solidities of $\epsilon_0 = 0.9987, 0.985, 0.962$, and $0.886$.
		A radial three-spoke pattern is used for the $m = 1$ layer porosity membranes and six-spoke patterns for membranes with $m = 3, 5$ and $9$ layers.
	}
	\label{fig:porous_patterns}
\end{figure*}%
The membranes in this study were created using a platinum-based addition-cure silicone rubber (\textit{Dragon Skin FX Pro, Shore Hardness 2A; Smooth-On Inc., Macungie, PA}).
This silicone rubber is prepared by mixing three components: Part A, a silicone hydride; Part B, a vinyl compound that functions as a catalyst for polymerization; and a thinning agent (\textit{TC 5005-C; BJB Enterprises Inc., Tustin, CA}), which adjusts the material stiffness of the cured membranes.
Membranes with three different thinning agent portions ($5\%$, $25\%$, and $45\%$) were fabricated, corresponding to shear moduli of $G = \left[\SI{32.04}{\kilo\pascal}, \SI{13.22}{\kilo\pascal}, \SI{6.53}{\kilo\pascal}\right]$, and locking parameters of $J_m = \left[28.97, 36.47, 47.52\right]$.
These materials were characterized and modeled in a previous study by \citet{das_nonlinear_2020}.
The membranes are then cast onto level plates and their thickness is controlled with a mechanical micrometer film applicator.
After the membranes are cured, the thickness of each membrane is measured with an outside micrometer.
The average thickness and standard deviation of all membranes used in this study was $h = 125 \pm \SI{5}{\micro\meter}$.

The porosity is implemented by cutting a symmetric pattern with prescribed pore diameters of $d_{\textnormal{set}} = \SI{0.61}{\milli\meter}$ onto the membrane sheets with a laser cutter.
The size of the pores is confirmed by measuring the pore diameters on the membranes in their zero-pressure state using the stereo camera setup (\Cref{fig:setupOverview}b) before each experiment; the average diameter and standard deviation across all samples was found to be $d = \SI{0.631}{\milli\meter} \pm \SI{0.049}{\milli\meter}$.
The patterns are based on a 6-spoke design— a symmetric, radially expanding layout that ensures consistent pore spacing and robust upscaling (\Cref{fig:porous_patterns}).
Starting at an initial radius $r_1$, six pores are distributed evenly around the circle with an angular spacing of $\Delta \phi = \ang{60}$.
Then, a second layer at a radius $r_2$ is chosen.
Pores are distributed on the same initial six angles $\phi$; in addition six more pores are placed, one between each spoke.
On the third layer, the process is repeated, but two additional pores are placed equidistantly between each spoke.
This pattern continues outward until the final $m$-th layer at radius $r_m$.
No pore is placed at the center of the pattern to enable for laser distance measurements of the membrane center-line.
The number of all pores can thus be calculated as $n = \sum_1^m 6 (m-1)$ for the different patterns.
The radii of each layer are chosen so that the difference between the minimal distance between two pores on a single layer and the distance between pores of adjacent layers (for non-spoke pores) is minimized ($\approx const.$) (\Cref{fig:porous_patterns}).
The initial radius for each layer is chosen relative to the diameter of the membrane $D$ at $r_m = 0.80 (D/2)$ for every pattern and expanded inwards until $r_1$.
Note that the only exception to this arrangement is the membrane with a single layer, which has three pores at $r_1 = r_m = 0.34 (D/2)$, positioned midway between the first and last layer of the multi-layer patterns.
The one-layer pattern was chosen as a minimal pore pattern while conserving the axisymmetric nature of the spoke patterns.
A total of 15 membranes were manufactured and tested, varying over three different material parameters ($G$ and $J_m$) and five different porosity patterns, including one non-porous membrane for reference.
The solidity is defined as the ratio between the closed area of the membrane over the membrane disc area $\epsilon = 1 - n \pi (d/2)^2$.
The initial, zero-stretch solidities range from $\epsilon_0 = 0.9987, 0.985, 0.962$, and $0.886$ with increasing layer number.
Throughout this work, porosity refers to the open area fraction, defined as $1 - \epsilon$.
\subsection{Pressure chamber}
\begin{figure*}
	\centering
	\includegraphics[width=0.90\textwidth]{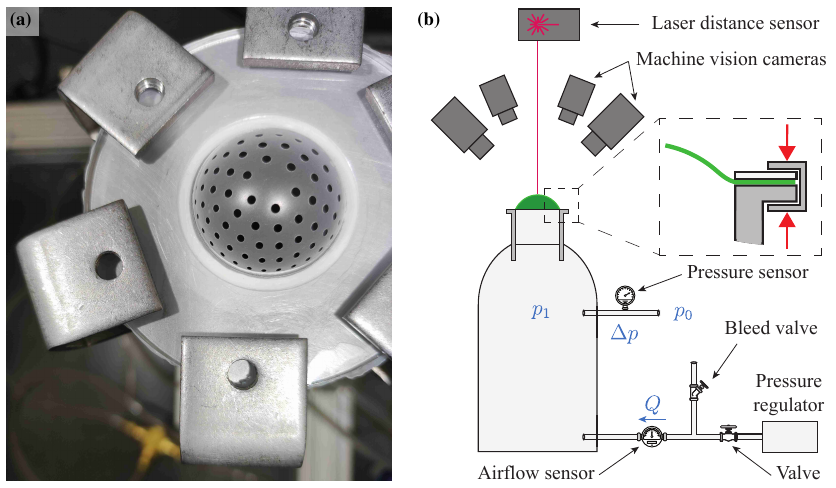}
	\caption[]{
		(a)~Top view on the experimental apparatus showing expanding pores for a five-layer pore pattern.
		(b)~Bulge test stand with direct measurements of pressure difference $\Delta p$, flow rate $Q$, and sample height $w_0$.
		Machine vision cameras in stereo configuration track the shape and local stretch of the samples.
	}
	\label{fig:setupOverview}
\end{figure*}
\begin{figure*}
	\centering
	\includegraphics{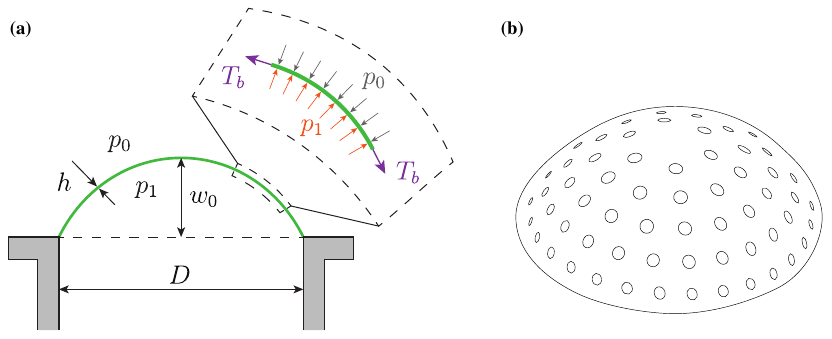}
	\caption[]{
		(a)~Dimensions of the expanding membrane and the relation between the internal and external pressure with the membrane tension,
		(b)~Schematic of expanding pores for a five-layer porosity pattern.
	}
	\label{fig:membraneModelSketch}
\end{figure*}

%
To characterize the material behavior of the poro-elastic membranes, different membrane samples are subjected to increasing pressure loading on a bulge test stand (\Cref{fig:setupOverview}b).
The pressure chamber consists of a large tank ($\approx \SI{1.5}{\liter}$) to ensure that the flow velocities inside the chamber are negligible, allowing the tank to act as a reservoir of effectively infinite size.
On top of the tank, a circular pipe with a flat flange is connected to provide an interface for mounting and testing the membrane samples.
The interface opening and effective diameter of the membranes is $D = \SI{0.03}{\meter}$ (\Cref{fig:membraneModelSketch}a).
The samples are mounted on an acrylic ring and secured in place with a thin plastic sheet.
The acrylic ring with the mounted samples is then positioned and sealed onto the pressure chamber interface (\Cref{fig:setupOverview}b, inset) using screw clamps.
The size of the samples versus the acrylic rings are chosen such that a desired pre-stretch ($\lambda_0$) is achieved when their outer diameters are aligned.
Measuring the pre-stretch before each test series ensures that the target pre-stretch is maintained after clamping.
We compare the centers of the pores from the imposed pattern (CAD model) with those measured using the stereo-camera setup.
The tank is supplied with a constant flow rate $Q$ at a pressure $p_1$ from an external pressure supply and is regulated by a mechanical valve.
In addition to the main valve, a bleed valve is installed to prevent pressure build-up in the chamber for non-porous and high-solidity samples.

The pressure difference $\Delta p$ between the interior $p_1$ and exterior $p_0$ is measured with a differential pressure transducer (\textit{Setra 26512R5WD2BT1F, Setra Systems Inc, Boxborough, Massachusetts}) located near the membrane, in the upper part of the tank.
The flow rate $Q$ passing through the porous membranes is determined by measuring the supply flow rate at the bottom of the tank using a digital air flow sensor (\textit{Honeywell Zephyr, HAFUHT0100L4AXT, Honeywell International Inc., Charlotte, NC}) rated up to $Q = \SI{1.67e-3}{\metre\cubed\per\second}$ (= 100 SLPM).
The deformation of the membrane center is measured with a laser distance sensor (\textit{AR700-16 Acuity Laser Sensor, AP7420160, Schmitt Measurement Systems, Portland, OR}) positioned above the membrane (\Cref{fig:setupOverview}b).
Four machine vision cameras (\textit{Alvium 1800 U-500m, Allied Vision, Stadtroda, Germany}) are positioned with equal spacing at $\ang{45}$ inclination relative to the membrane centerline to track the position of the pores and additional tracking markers on the samples.
The stereo camera tracking system allows us to reconstruct the full 3D shape of the porous membranes, including the local stress and strain at each position on the membrane.
Additionally, we conduct direct pore size measurements for a few select cases by tracking the outline of the pores in 3D space, which allows us to relate the stretch to the pore size directly.

All measurement devices are synchronized, and data is recorded over five seconds after waiting 30 seconds for the pressure to equilibrate.
The recorded signals are then time-averaged for further processing and analysis.
One set of stereo images is taken at each pressure level $\Delta p$.
The entire system is considered stationary, as no discernible time variance of $\Delta p$, $Q$, or $w_0$ is observed at a fixed measurement point where $\Delta p$ is regulated by the valve position.
\section{Results and Discussion}
In this study, we characterize the behavior of poro-elastic membranes under various pressure loadings and establish scaling laws relevant to the design of poro-elastic engineering applications.
We begin by analyzing the deformation of the membranes in response to applied pressure, followed by an examination of the relationship between membrane deformation and through-flow.
\subsection{Elastic characterization and modeling} \label{sec:elastic_characterization}
\begin{figure}
	\centerline{\includegraphics{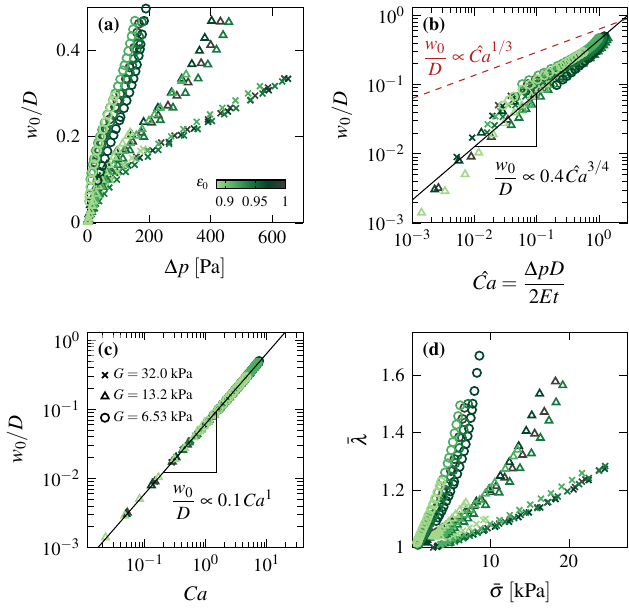}}
	\caption{
		(a) Normalized membrane center-line deformation $w_0 / D$ as a function of the pressure difference $\Delta p$ acting on the membrane,
		(b) Normalized membrane center-line deformation as a function of the Cauchy number $\hat{\Ca}$ based on linear membrane theory,
		(c) Normalized membrane center-line deformation as a function of the Cauchy number \Ca from the Gent model, and
		(d) Average membrane stretch $\bar{\lambda}$ as a function of average membrane stress $\bar{\sigma}$, shown for different solidities $\epsilon_0$ and materials $G$.
	}
	\label{fig:centerLineDeformation}
\end{figure}%
Initially, the membranes are flat disks with a diameter $D$ and a pre-stretch $\lambda_0$ applied during mounting.
The pre-stretch is measured individually before each experiment by comparing the pore positions with the unstretched radial pattern prior to mounting.
An additional zero-pressure measurement is performed after each measurement series to ensure that the membranes have not slipped from their mounts.
The membrane sample diameter and mounting ring diameters were carefully selected such that aligning them set the pre-stretch to a target value (see also the supplementary material of \citep{das_nonlinear_2020}).
For our experiments, the target was 1\% pre-stretch, which was achieved in the majority of cases with an average and standard deviation of $\bar{\lambda}_0 = 1.014 \pm 0.022$.
Notably, one case with a 6.5\% pre-stretch appears as an outlier in later parts of the analysis (minimal porosity $\epsilon_0$ with the lowest material stiffness $G$).

When the pressure difference ($\Delta p$) between the inner chamber and the ambient pressure increases, the membranes bulge outward, forming a spherical cap with a maximum height ($w_0$) at the center for both porous and non-porous membranes (\Cref{fig:membraneModelSketch}a).
This spherical cap deformation has been previously demonstrated by \citet{flory_deformation_2007} and \citet{das_nonlinear_2020}, and is also observed in our experiments with porous membranes.

To develop a general expression for the membrane-average stress ($\bar{\sigma}$) and stretch ($\bar{\lambda}$), we measure the center-line deformation ($w_0$) and pressure difference ($\Delta p$) across a range of material properties and porosity patterns (\Cref{fig:centerLineDeformation}a).
Note that variables with an overbar $(\bar{\phantom{u}})$ represent membrane-averaged quantities.
In the following analysis, we also examine the local stretch as a function of radial position and applied loading, $\lambda(r,\sigma)$, as detailed in \Cref{sec:local_stretch}, and compare it against the membrane-averaged quantities.

The center-line deformation reaches up to $w_0/D = 0.5$ for the most compliant membranes.
The applied pressure varies from $\Delta p = 0$ to $\SI{650}{\pascal}$.
The membrane deformation groups by stiffness, indicated by different markers, while porosity has minimal impact on the deformation variation, as shown by the different colors.
The silicone-based elastomers exhibit hyper-elastic behavior under pressure loading, consistent with the findings of \citet{das_nonlinear_2020}.
Hyper-elastic materials can sustain large strains before strain-stiffening occurs, as seen in the inflection points for intermediate and highly flexible membranes (\Cref{fig:centerLineDeformation}a).

To further contextualize the membrane deformation, we compare our data to the classical axisymmetric linear elastic membrane scaling of the form,
\begin{equation}
	\frac{w_0}{D} \propto \hat{\Ca}^{1/3} ,
\end{equation}
where $\hat{\Ca} = \Delta p D / (2 E h)$ is a Cauchy number based on the linear response of the membrane with an Elastic modulus $E$.

As shown in \Cref{fig:centerLineDeformation}b, our data systematically deviates from this prediction.
However, we find that the data groups around a modified empirical scaling of the form
\begin{equation}
	\frac{w_0}{D} = 0.4 \hat{\Ca}^{3/4}
\end{equation}
across three orders of magnitude in both $w_0 / D$ and $\hat{\Ca}$.
Note that this relationship is sometimes expressed with $2 w_0 / D = \alpha \hat{\Ca}^\gamma$ in the literature which would then yield a pre-factor of $\sim 0.8$ instead.
While this simplified model does a decent job in predicting the deformation-pressure relationship, it has a fair amount of spread, especially around $\hat{\Ca} \approx 0.1$.
This suggests that the deformation response is governed by nonlinear effects not captured by linear elastic theory.
The deviation motivates us to use of a hyperelastic constitutive model (Gent model), while the existence of a power-law trend indicates an underlying self-similar mechanical response \citep{long_axisymmetric_2012}.

Past studies have demonstrated that the biaxial stress–strain relationship of hyperelastic membranes is well described by a two-parameter Gent model \citep{das_nonlinear_2020}:
\begin{equation}
	\sigma = G_m \left(\bar{\lambda} - \frac{1}{\bar{\lambda}^2}\right),
	\label{eq:gent_model}
\end{equation}
where the effective shear modulus is given by the model
\begin{equation}
	G_m = \frac{G J_m}{J_m - I_1 + 3}.
\end{equation}
Here, $G$ is the shear modulus, and $J_m$ is the locking parameter, which characterize the Neo-Hookean response of the material, and $I_1$ is the first invariant of the deformation tensor.
The material parameters can be determined either from uniaxial tensile tests or directly from bulge test measurements.
For our experiments we draw on the material characterization done in a past study on a similar experimental setup in our lab \citep{das_nonlinear_2020}.

Local deformation analysis (\Cref{sec:local_stretch}) confirms that the biaxial stress-strain relationship provides a good approximation of the overall membrane deformation.
However, its accuracy diminishes for large deformations ($w_0 \rightarrow 0.5 D$) and near the membrane mount ($r_0 \rightarrow 0.5 D$), where additional effects may become significant.

With the stress and stretch the membrane tension ($T_b$) can be calculated from the first Piola-Kirchhoff stress according to \citet{das_nonlinear_2020}:
\begin{equation}
	T_b = G_m h \left(1 - \frac{1}{\bar{\lambda}^6}\right) \, .
\end{equation}%
We introduce a Cauchy number as the ratio of membrane tension to pressure loading:
\begin{equation}
	\Ca = \frac{\Delta p D}{T_b} \, .
\end{equation}%
The Cauchy number correlates well with the membrane's center-line deformation (\Cref{fig:centerLineDeformation}c), and the data collapses on one line with a power law dependence with a slope of $w_0/D \propto \Ca^1$.
Importantly, for these relatively low levels of porosity (solidity ranges from $\epsilon_0 = 0.886$ to $0.9987$) the porosity does not have a discernible effect on the structural response of the membrane, as its relation with any applied stress is consistent between patterns.
This was initially not expected when we conceptualized this study and we expect this to eventually not hold anymore at higher levels of porosity not covered in this study.

This scaling relationship allows us to normalize and predict membrane deformation as a function of pressure if the material properties are known or determined via the bulge tests as done in this study.
Furthermore, these predictions can be applied to systems where pressure is applied by fluid dynamic loading ($q = 0.5 \rho U^2$), such as in experiments with dynamic loading \citep{pezzulla_deformation_2020, marzin_flow-induced_2022, mathai_shape-morphing_2023}.

The elastic scaling model allows us to compare internal membrane stress relative to stretch (\Cref{fig:centerLineDeformation}d).
The membrane-average stretch  $\bar{\lambda}$ is determined through the Gent model (\Cref{eq:gent_model}) which includes the pre-stretch $\bar{\lambda}_0$ determined under zero pressure conditions at the beginning of the experiments.
The non-linear behavior observed in the stress-stretch curves is well captured by the model, displaying consistent curvature for each of the three different materials which agrees well with other studies on hyper-elastic membranes \citep{sasso_characterization_2008, das_nonlinear_2020}.
\subsection{Local membrane stretch} \label{sec:local_stretch}
\begin{figure*}
	\centering
	\includegraphics[width=0.40\textwidth]{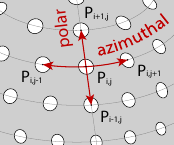}
	\caption[]{
		Local stretch $\lambda_{i,j}$ definition for each radial position of pores.
	}
	\label{fig:local_stretch_sketch}
\end{figure*}%
\begin{figure*}
	\centerline{\includegraphics{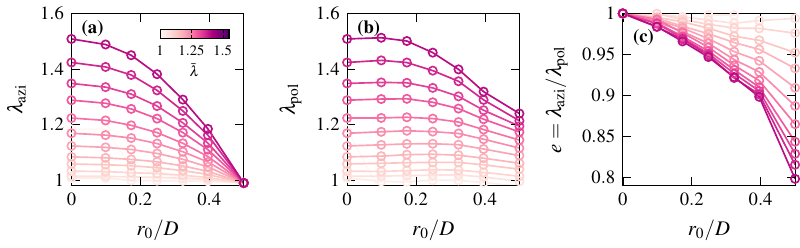}}
	\caption{
		(a) Azimuthal stretch ($\lambda_{\textnormal{azi}}$),
		(b) Polar stretch ($\lambda_{\textnormal{pol}}$), and
		(c) Stretch eccentricity ($e = \lambda_{\textnormal{azi}} / \lambda_{\textnormal{pol}}$) as a function of initial radial position ($r_0$),
		for different radial-average stretches ($\lambda$) indicated by the color gradient.
		The local stretch is shown for a five-layer pattern and low stiffness ($G = \SI{6.5}{\kilo\pascal}$), but is representative for all tested membranes.
	}
	\label{fig:localStretchEcc}
\end{figure*}%
In order to understand the flow through the porous membrane, we need to first determine the geometry of the evolving pore distribution of the expanding membranes.
We measure the pore positions using the pores itself as tracers, and applying markers to membranes without or only very few pores.
The markers are applied by placing small paint drops on the membrane that do not penetrate into the silicon rubber.
The positions of all pores are tracked optically (\Cref{fig:setupOverview}b) to determine the local stretch $\lambda(r,\bar{\lambda})$ for each pore based on its radial position and the overall deformation of the membrane.
The azimuthal and polar stretches are calculated on the spherical cap membrane shape using the relative positions of the pores (\Cref{fig:local_stretch_sketch}).
The azimuthal stretch ($\lambda_{\textnormal{azi}}$) is determined by the relative distance along the membrane surface between two adjacent pores at the same radial position, while the polar stretch ($\lambda_{\textnormal{pol}}$) is determined by the relative radial distance between pores at adjacent radial positions:
\begin{equation}
	\lambda_{\textnormal{azi}}(r,\lambda) = l_{\textnormal{azi}}(r,\lambda) - l_{\textnormal{azi},0}(r) \quad \textnormal{and} \quad \lambda_{\textnormal{pol}}(r,\lambda) = l_{\textnormal{pol}}(r,\lambda) - l_{\textnormal{pol},0}(r) \; ,
\end{equation}%
with central differences $l_{\textnormal{azi}}(\lambda) = 0.5 \; \overline{P_{i,j-1} P_{i,j+1}}$, and $l_{\textnormal{pol}}(\lambda) = 0.5 \; \overline{P_{i-1,j} P_{i+1,j}}$ (\Cref{fig:local_stretch_sketch}).

Due to the symmetry of the pore pattern, the local stretch remains constant between different pores at the same radial position ($r$).
The azimuthal stretch increases across all radial positions with increasing deformation, except at the position closest the membrane edge (\Cref{fig:localStretchEcc}a), where the membrane is clamped and cannot expand azimuthally.
Note that the local stretch data is plotted here as a function of the initial radial position $r_0$ for visual clarity whereas in the bulging membrane the radial position of the pores ($r$) changes with the stretch accordingly.
Here, the azimuthal stretch remains at the pre-stretch value $\lambda_0$.
With higher average stretch ($\lambda$), the azimuthal stretch ($\lambda_{\textnormal{azi}}$) is greater at the center compared to the outer parts of the membrane where the membrane is geometrically restricted from expanding.
The radial stretch evolution could also be observed on hyper-elastic materials tested on a bulge setup by \citet{sasso_characterization_2008}, although its implications for their study are not further discussed.
The polar stretch ($\lambda_{\textnormal{pol}}$) initially stretches uniformly across the membrane (\Cref{fig:localStretchEcc}b) but shows higher stretch near the membrane's center at higher deformations as well.

The difference between azimuthal and polar stretches results in a stretch eccentricity ($e$), which varies with the initial radial position ($r_0$) and the membrane's overall deformation given by $\bar{\lambda}$.
Near the membrane's center, the stretch remains bi-axial.
Towards the outer regions, the stretch becomes more eccentric with increased $\bar{\lambda}$ due to restricted azimuthal stretch at the mounting radius ($r_0 = 0.5 D$).
The eccentricity reaches up to $e = 0.8$ for the highest deformation presented ($\bar{\lambda} = 1.5$).
These results indicate that the assumption of biaxial stress-strain holds across most of the membrane, except at the highest stretches and near the membrane mount.

\subsection{Pore size evolution} \label{sec:pore_size_evolution}

\begin{figure*}
	\centerline{\includegraphics[]{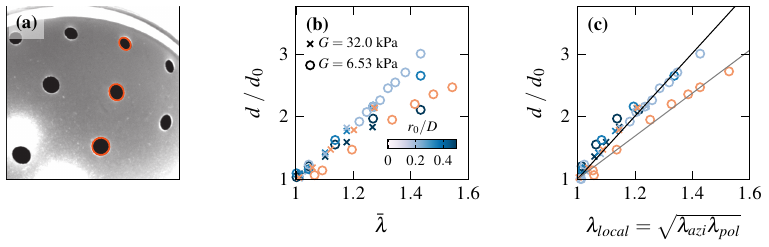}}
	\caption{
		(a) Close-up view of directly measured pores (red lines) from one of the stereo-imaging cameras,
		(b) Directly measured pore diameter $d$ as a function of the membrane-average stretch $\bar{\lambda}$,
		(c) Directly measured pore diameter $d$ as a function of the local stretch $\lambda(r,\bar{\lambda})$.
		Plots shown for two three-layer membranes in blue ($\epsilon_0 = 0.985$) and two one-layer membranes in orange ($\epsilon_0 = 0.9987$).
		The markers indicate different material stiffness, and the color shading indicates the initial radial pore size position for the multi-layer membranes.
		The solid lines represent the best linear fit to the data.
	}
	\label{fig:directPores}
\end{figure*}%

Here we describe how the pore diameter scales with local ($\lambda$) and membrane-average stretch ($\bar{\lambda}$).
We aim to establish a predictive relationship for the pore size based on observable parameters, specifically pressure and centerline deformation, both of which relate to $\bar{\lambda}$ as shown in \Cref{sec:elastic_characterization}.

In addition to the local stretch measurements discussed in \cref{sec:local_stretch}, we directly measure the pore size for four different membranes using the stereo-deformation images (\Cref{fig:directPores}a).
The normalized pore diameter, $d/d_0$, is shown for two one-layer and two three-layer membranes with different stiffnesses in \Cref{fig:directPores}b,c.
Markers indicate different material stiffnesses, while colors represent the pore’s initial radial position $r_0$ on the membrane (see \Cref{fig:porous_patterns}).
Throughout the rest of the paper, the pore diameter is determined from the pore area as $d = 2 \sqrt{A/\pi}$, while we note that pore eccentricity can reach values $e > 0.8$ at higher $\lambda$ (\cref{sec:local_stretch}).

We observe that pores expand faster than both the membrane-average and local stretch (\Cref{fig:directPores}b,c), reaching over three times their initial diameter $d_0$.
Similar hole expansion behavior has been demonstrated by \cite{cohen_cavitation_2010} for hyperelastic materials (Poisson ratio $\nu = 0.5$) with a single pore at the center.
At low stretch values ($\bar{\lambda} < 1.2$), the pore diameter increases approximately linearly with $\bar{\lambda}$ across all radial positions and membranes.
At higher stretches, pores near the membrane edge grow at a slower rate.

The non-uniform pore expansion is caused by local stretches becoming more concentrated near the center of the membrane at large $\bar{\lambda}$ values (\Cref{fig:localStretchEcc}b,c).
Interestingly, the pore diameter shows a consistent linear dependence on the local stretch $\lambda$ (\Cref{fig:directPores}c) for the majority of the pores measured.
The only deviation from that trend is found for the soft, single-layer pores ($G = \SI{6.53}{\kilo\pascal}$) from the membrane with a higher applied pre-stretch $\lambda_0$ (\Cref{sec:method_membranes}).
We use this linear relationship to compute the pore diameter $d/d_0$ in all cases where local stretch measurements are available.

\begin{figure*}
	\centerline{\includegraphics{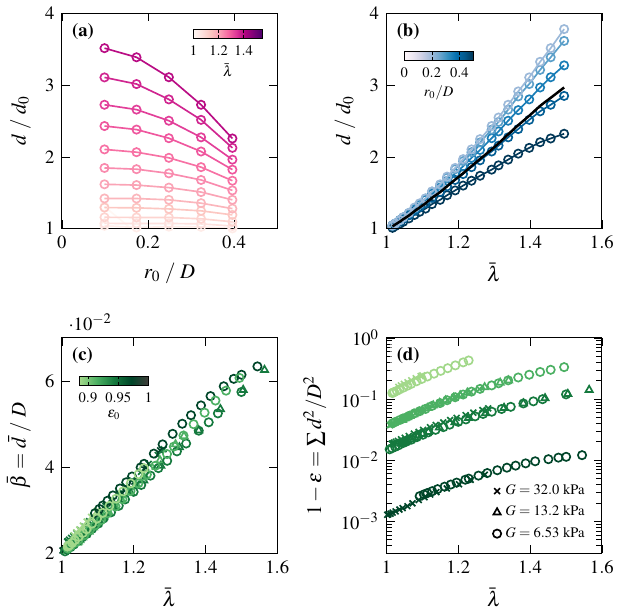}}
	\caption{
		(a) Local pore diameter* $d / d_0$ as a function of the initial radial position $r_0/D$ for various membrane-average stretches $\bar{\lambda}$.
		(b) Local pore diameter as a function of $\bar{\lambda}$; colors correspond to initial radial position $r_0$.
		The black line represents the membrane-average diameter $\bar{d}$.
		(c) Membrane-average diameter ratio $\bar{\beta}$ versus membrane-average stretch $\bar{\lambda}$.
		(d) Porosity $1 - \epsilon$ versus membrane-average stretch $\bar{\lambda}$;
		colors indicate the initial solidity $\epsilon_0$.\\
		*Local diameters in (a) and (b) are shown for a five-layer, low-stiffness membrane ($G = \SI{6.5}{\kilo\pascal}$), but are qualitatively and quantitatively representative of all tested membranes.
	}
	\label{fig:localPoreDiameter}
\end{figure*}%

Using the scaling between local stretch and pore diameter, we determine pore size evolution as a function of $\bar{\lambda}$ and radial position $r$ for all membranes (\Cref{fig:localPoreDiameter}).
In \Cref{fig:localPoreDiameter}a,b the pore diameter as a function of the radial position $r_0/D$ is presented for one membrane (five-layers at $G = \SI{13.2}{\kilo\pascal}$) before generalizing to all porosity levels and stiffness values in \Cref{fig:localPoreDiameter}c,d.

At small $\bar{\lambda}$, pore diameters remain uniform across $r_0/D$, with $d \approx d_0$.
As $\bar{\lambda}$ increases, central pores ($r_0 \approx 0.1 D$) expand up to $d = 3.5 d_0$.
A radial gradient emerges, with decreasing pore size toward the membrane edge, reflecting the local stretch distributions in \Cref{fig:localStretchEcc}.
This gradient intensifies with larger deformation.
At $\bar{\lambda} = 1.6$, the diameter difference between central and edge pores exceeds 35\%.

To quantify the pore size evolution over all membranes, we define the membrane-average diameter $\bar{d}$ from the pore-average area $\bar{a}$:
\begin{equation}
	\bar{d} = \sqrt{\frac{4}{\pi n}\sum^{n} a(r)}
\end{equation}%
This parameter characterizes the global pore evolution and will be linked to flow rate predictions in later sections.
We find that $\bar{d}$ lies approximately midway between the minimum and maximum local diameters (\Cref{fig:localPoreDiameter}b).
The average diameter increases approximately linearly with $\bar{\lambda}$.
Pores near the center ($r_0 < 0.3 D$) grow faster than average, while those near the edge ($r_0 > 0.3 D$) expand more slowly.

Comparing different porous patterns, the membrane-average diameter ratio $\bar{\beta} = \bar{d}/D$ collapses onto a single linear trend for all membranes (\Cref{fig:localPoreDiameter}c).
This suggests self-similarity among the different patterns, when the pore diameter is related to the local and membrane-average stretches.
The diameter ratio is $\bar{\beta} = 0.021$ for the unstretched membranes and grows linearly to more than three times the initial value ($\bar{\beta} = 0.063$) at the highest membrane-average stretch.

Lastly, we examine porosity ($1 - \epsilon$) as a function of $\bar{\lambda}$ in \Cref{fig:localPoreDiameter}d.
We compute porosity by summing total pore area and normalizing by the membrane area: $\sum d^2 / D^2$, or equivalently $n \bar{d}^2 / D^2$.
Porosity varies across patterns, as expected, but shows consistent scaling with membrane stiffness and shows the robustness of the measurements over three orders of magnitude.
This validates that the same pore pattern will yield consistent porosity for a given $\bar{\lambda}$.
Although initial porosities are low ($1 - \epsilon_0 = 0.1\%, 1.5\%, 3.8\%$, and $11.4\%$ for increasing layer number), porosity increases by more than a factor of almost ten at $\bar{\lambda} = 1.6$.

We summarize the key findings that emerge from the local stretch and pore evolution analysis:
First, local stretch is highest near the membrane center, leading to larger pore diameters and open areas in that region.
At large global deformations, pore stretch becomes increasingly eccentric at the outer radii due to differences between azimuthal and polar stretch.
Second, the pore diameter scales linearly with local stretch.
Third, the membrane-average diameter ratio scales linearly with $\bar{\lambda}$, and the data collapse well across all patterns and materials.
This leads to membrane-average diameter ratio increasing by a factor of more than three and consequently porosities increasing by a factor of almost ten.
In the following sections, we use the resulting pore area evolution to predict flow rates through the porous membranes.

This non-uniform pore expansion may be leveraged for flow control or filtration applications, in a manner analogous to how kirigami sheets promote three-dimensional, spatially varying geometries \citep{marzin_flow-induced_2022}.
The resulting porosity gradients could influence the characteristic length scales of turbulent structures downstream.
Moreover, introducing multiple length scales into a turbulent flow is known to enhance mixing efficiency and could enable applications at high Reynolds numbers.
\subsection{Flow rate analysis}
\begin{figure*}
	\centerline{\includegraphics[]{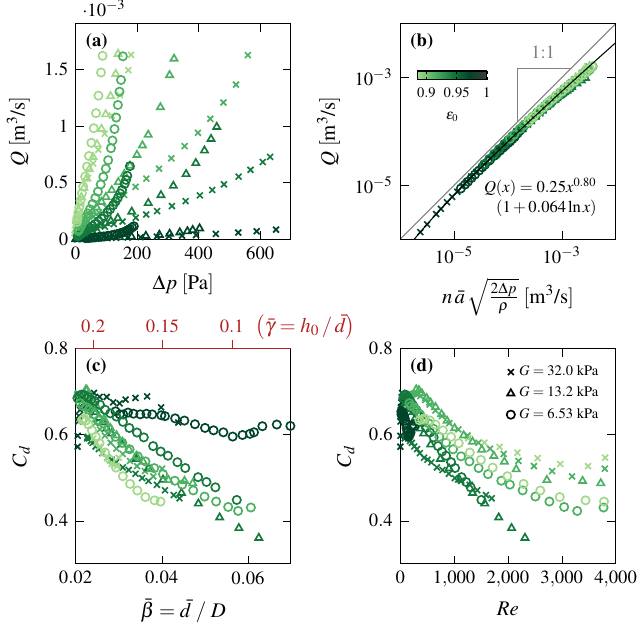}}
	\caption{
		(a) Measured volume flow rate $Q$ as a function of the pressure difference $\Delta p$,
		(b) Measured volume flow rate $Q$ versus ideal flow rate driven by the pressure difference $\Delta p$,
		(c) Discharge coefficient $C_d$ as a function of the membrane-average diameter ratio $\bar{\beta}$ and thickness-to-diameter ratio \textcolor{groseille}{$\bar{\gamma}$},
		(d) Discharge coefficient $C_d$ as a function of the pipe Reynolds number \textit{Re}.
		All plots are shown for different initial solidities ($\epsilon_0$-color gradient) and material stiffnesses ($G$-markers) as indicated in the legends.
	}
	\label{fig:flowRate}
\end{figure*}
We now examine the relationship between the measured flow rate $Q$ through the membrane pores and the applied pressure difference $\Delta p$.  
The flow rate reaches up to $Q = \SI{1.62e-3}{\metre\cubed\per\second}$ at the highest pressures ($\Delta p \approx \SI{650}{\pascal}$), particularly for membranes with intermediate stiffness and initial solidity (\Cref{fig:flowRate}a).  
Softer membranes achieve the highest flow rates at relatively low pressures, whereas stiffer membranes with low initial porosity are restricted to lower flow rates before reaching either the sensor limit or the maximum considered membrane expansion ($w_0 = 0.5 D$).

Because the membrane-average stretch $\bar{\lambda}$ varies with pressure, a single flow rate $Q$ can correspond to different porous geometries—either due to differences in initial solidity or to variations in the resulting average pore area at a given deformation. Conversely, membranes with apparently similar geometries can produce different flow rates, depending on their mechanical response.

Understanding how flow rate scales with membrane stiffness and pressure loading requires considering both the pressure differential $\Delta p$, which drives the flow, and the expanded solidity $\epsilon$ (\Cref{fig:localPoreDiameter}d), which governs the available flow area through the membrane.

The discharge coefficient ($C_d$) accounts for losses due to orifice geometry and is defined as the ratio of the measured flow rate to the ideal flow rate through an infinitely large orifice \citep{merritt_hydraulic_1967, miller_flow_1996}:
\begin{equation}
	C_{d,\textnormal{single}} = \frac{Q_{\textnormal{exp}}}{Q_{\textnormal{ideal}}} = \frac{Q}{a\sqrt{\frac{2 \Delta p}{\rho}}} \; ,
\end{equation}
where $a$ is the orifice or pore area and $\rho$ is the fluid density.

Treating the membrane as a collection of non-interacting orifices in parallel, we define a membrane-averaged discharge coefficient:
\begin{equation}
	C_d = \frac{Q_{\textnormal{exp}}}{Q_{\textnormal{ideal}}} = \frac{Q}{n \bar{a}\sqrt{\frac{2 \Delta p}{\rho}}} \; ,
	\label{eq:discharge_coefficient}
\end{equation}
where $n$ is the number of pores and $\bar{a}$ is the average pore area, calculated from pore deformation at different pressures (see \Cref{sec:local_stretch}).

We compare these two contributions by plotting the measured flow rate versus the ideal flow rate (\Cref{fig:flowRate}b), using $\bar{a}$ determined from the pore size evolution in the previous section.  
The ratio remains consistent for all tested membranes across nearly four decades of flow rate.  
All data points fall to the right of the ideal line, indicating a pressure drop across the membrane.  
Although the curve does not follow a strict power law, it captures the overall behavior across several orders of magnitude and multiple measurement systems (flow meter, pressure sensor, deformation tracking), underscoring the robustness of the analysis.  
The flow rate can be predicted empirically using a modified power law of the form $Q(x) = A x^B (1 + C \ln x)$, where $x$ is the ideal flow rate (\Cref{eq:discharge_coefficient}).

From a dimensional standpoint, the discharge coefficient can depend on the following parameters \citep{kolodzie_discharge_1957, huang_study_2013}:

\begin{itemize}
	\item Fluid velocity through the pores ($u$)
	\item Fluid density ($\rho$)
	\item Dynamic viscosity ($\mu$)
	\item Pore diameter ($d$)
	\item Plate thickness ($h$)
	\item Pitch angle ($\theta$)
	\item Plate area ($A$)
	\item Number of pores ($n$)
\end{itemize}

For the poro-elastic system presented in this study, all but the last two parameters vary from one experiment to the next.  
This variability makes it challenging to control the relevant non-dimensional groups.

We begin by presenting the discharge coefficient as a function of two geometric factors: the diameter ratio $\bar{\beta} = \bar{d} / D$ and the inversely proportional thickness-to-diameter ratio $\bar{\gamma} = h_0 / \bar{d}$ (\Cref{fig:flowRate}c).  
The thickness-to-diameter ratio is presented relative to the initial membrane thickness $h_0$, since no direct measurements of the thickness in the deformed state are available, and the non-uniform stretch leads to complex local membrane thinning \citep{cohen_cavitation_2010}.

A simple volume-conservation argument between the undeformed disk and the hemispherical shape allows us to estimate that the membrane-average thickness could be reduced to about $0.5 h_0$ for an incompressible material with initial volume $V_0$ ($= 0.25 \pi D^2 h_0$):
\begin{equation}
	h(w_0) = h_0 \frac{D^2}{D^2 + 4w_0^2} \, .
	\label{eq:thickness_model}
\end{equation}
The discharge coefficient starts out between $C_d = 0.6$ and $0.7$ and gradually falls to $C_d = 0.4$ with increasing $\bar{\beta}$ or decreasing $\bar{\gamma}$ for the multi-layer membranes.  
The single-layer membranes maintain a discharge coefficient close to $C_d \approx 0.65$ across the $\bar{\beta}$ range.  
A decreasing discharge coefficient with increasing diameter ratio or decreasing thickness-to-diameter ratio has also been reported by \cite{kolodzie_discharge_1957} and \cite{huang_study_2013}.  
In their experiments, \cite{huang_study_2013} reported $C_d$ values decreasing to as low as $0.65$ for minimum $\bar{\gamma} = 0.1$, but much higher $\beta = 0.6$.  
In our experiments, the diameter ratio and thickness ratio are nearly an order of magnitude lower, $\bar{\beta} = 0.07$ and $\bar{\gamma} = 0.057$ (or $\bar{\gamma} = 0.029$ using \Cref{eq:thickness_model}), which may explain the lower discharge coefficients observed.

These trends in $C_d$ must also be interpreted alongside other non-dimensional parameters governing the system, particularly the fluid velocity through the pores $u$, the fluid density $\rho$, and the fluid’s dynamic viscosity $\mu$.  
To assess the role of viscous versus inertial effects, we define a pipe Reynolds number (\Rey) using the plate diameter $D$ and pipe flow velocity ($U = Q / (0.25 \pi D^2)$):
\begin{equation} \label{eqn:reynolds_number2}
	\Rey = \frac{\rho U D}{\mu} = \frac{4 \rho Q}{\pi \mu D}
\end{equation}
Here, viscosity and density are evaluated for air at a temperature of $\SI{20}{\degreeCelsius}$, corresponding to the ambient lab temperature measured next to the experimental setup.

In our experiments, $\Rey$ ranges from 10 to 4000 depending on membrane stiffness and porosity (\Cref{fig:flowRate}d).  
The discharge coefficient falls from $C_d = 0.7$ to $0.4$ across all tested membranes and begins to level out at $\Rey > 3000$.  
These \Rey values fall within the transitional range and reflect shifting contributions from viscous and inertial forces, making $C_d$ a highly nonlinear function of flow conditions \citep{kolodzie_discharge_1957, merritt_hydraulic_1967, huang_study_2013}.  
While the discharge coefficient can initially rise with increasing \Rey at $\Rey < 10^3$, it eventually falls and asymptotically approaches a steady value for $\Rey > 10^4$.  
An especially sharp decrease in $C_d$ is found at low thickness-to-diameter ratios \citep{huang_study_2013}.  
The initial discharge coefficient of $C_d \approx 0.7$ is consistent with single-orifice results at low to intermediate Reynolds numbers \citep{merritt_hydraulic_1967, huang_study_2013}.

Similar to the results of $C_d$ as a function of $\bar{\beta}$ and $\bar{\gamma}$, the effects of \Rey in this study cannot be evaluated in isolation.  
While \Rey varies widely, $\bar{\beta}$ and $\bar{\gamma}$ also span similar orders of magnitude.  
Finally, the influence of the pitch angle $\theta$ is not explicitly included in this discussion.  
As the membranes expand into a hemisphere, pores further from the center develop an increasing pitch angle.  
At a hemispherical shape, membranes near the mount reach $\theta \approx \ang{90}$, while pores closer to the center remain at low $\theta$.

In summary, our flow rate analysis of the poro-elastic membranes demonstrates that several geometric and flow parameters are interdependent and cannot be fully decoupled.  
The discharge coefficient varies with the diameter ratio, the thickness-to-diameter ratio, and the Reynolds number.  
The flow through the porous membranes can be predicted from the ideal flow rate using a modified power law spanning more than three decades, but establishing a unified scaling for the discharge coefficient is left for future work.
\section{Conclusion}
This study provides a comprehensive experimental investigation of the coupled deformation and flow behavior of thin, compliant porous membranes under pressure loading. Prior work has primarily relied on theoretical or numerical analyses, with limited experimental data available. Here, we expand the empirical foundation by systematically exploring how material stiffness and porosity influence membrane mechanics and flow.

Using a controlled bulge test setup, we fabricated and tested silicone membranes with initial porosities ranging from solid ($\epsilon_0 = 1$) to moderately porous ($\epsilon_0 = 0.89$), and varying material stiffnesses. Our measurements of membrane deformation, local stretch, and flow rate reveal that porosity has a negligible influence on the membrane's deformation behavior, which is well captured by a two-parameter Gent model developed for hyper-elastic solids.

A key finding is that local membrane stretch increases toward the center, leading to non-uniform pore growth. Pore diameter scales linearly with local stretch, allowing us to relate the membrane-average pore diameter, and thus porosity, to either pressure loading or centerline deformation. Membranes were found to increase their porosity by up to a factor of ten when adopting a hemispherical shape.

To model flow behavior, we introduce a discharge coefficient that scales the measured flow rate to the ideal rate through the open pore area. This coefficient incorporates the evolving pore geometry and enables comparison across membranes.
Our results show that the discharge coefficient ($C_d$) decreases systematically with increasing pore diameter and Reynolds number, dropping from about $C_d = 0.7$ to $0.4$ as membranes transition from low-stretch to highly deformed states. Single-layer membranes maintain a relatively stable $C_d$, while multi-layer membranes exhibit sharper reductions, especially at low thickness-to-diameter ratios. The flow rate across all membranes can be predicted from the ideal open-pore flow using a modified power-law relation over more than three decades of flow rate, demonstrating the robustness of this scaling framework.

At the same time, the experiments reveal how closely coupled the system’s governing parameters are: pore geometry, membrane stretch, thickness, and Reynolds number all evolve together and cannot be varied independently. This interdependence complicates attempts to isolate single effects and points to the need for targeted experiments and modeling efforts that can disentangle these relationships.

Future studies should explore different levels of pre-stretch, controlled variations in porosity patterns, and simplified cases such as single poro-elastic orifices to clarify fundamental scaling laws. Numerical simulations that can access local pore-scale stresses and flow fields would complement experiments and help define where inertial, viscous, and geometric effects dominate. Such efforts will refine the predictive framework introduced here and expand its applicability.

Together, these results unify structural deformation and flow behavior into a predictive scaling framework, enabling the design of compliant porous systems tailored to specific performance needs. The experimental approach and scaling laws developed here have broad relevance for fluid–structure interaction systems, soft filtration technologies, and bio-inspired aerodynamic applications. The ability to modulate porosity and discharge behavior through geometry and material properties presents opportunities for adaptive flow control, directional filtering, and dynamic permeability in next-generation membrane systems.
\section*{CRediT authorship contribution statement}

\textbf{Alexander Gehrke:} Writing – original draft, Methodology, Investigation, Conceptualization, Formal analysis.
\textbf{Zoe King:} Writing – editing, Resources, Investigation, Visualization.
\textbf{Kenneth S. Breuer:} Writing – review \& editing, Supervision, Conceptualization, Funding acquisition.

\section*{Declaration of competing interest}

The authors declare that they have no known competing financial interests or personal relationships that could have appeared to influence the work reported in this paper.

\section*{Acknowledgments}

We sincerely thank the reviewers for their time and constructive feedback, which greatly improved the clarity and quality of the manuscript.
We also appreciate the editor's guidance throughout the review process.
This work was financially supported by the United States National Science Foundation (NSF) grant GR5260547.

\section*{Data availability}

Data will be made available on request.

\bibliographystyle{elsarticle-harv} 
\bibliography{literature}
\end{document}